\documentclass[9pt,conference]{IEEEtran}
\IEEEoverridecommandlockouts

\usepackage[utf8]{inputenc}
\usepackage{pgfplots}
\DeclareUnicodeCharacter{2212}{-} 
\usepgfplotslibrary{groupplots,dateplot}
\usetikzlibrary{patterns,shapes.arrows}
\pgfplotsset{compat=newest}

\usepackage{graphicx}
\usepackage{wrapfig}
\usepackage{multirow}
\usepackage{amssymb}
\usepackage{amsmath}
\usepackage{amsthm}
\usepackage{bm}
\usepackage{xfrac}
\usepackage{nicefrac}
\usepackage{enumitem,kantlipsum}
\usepackage{algorithm}
\usepackage{algpseudocode}
\usepackage{verbatim}
\usepackage{todonotes}
\usepackage{xurl}
\usepackage{slantsc}
\usepackage{booktabs}

\usepackage{subcaption}

\algtext*{EndFor}
\algtext*{EndIf}
\algtext*{EndWhile}

\usepackage{verbatim}
\usepackage{mathtools}
\usepackage{subfiles}
\usepackage{enumitem}
\usepackage[mathscr]{eucal}
\usepackage{enumitem,kantlipsum}
\usepackage{xspace}

\usepackage[most]{tcolorbox}

\newcommand\copyrighttext{%
  \footnotesize \textcopyright 2025 IEEE.  Personal use of this material is permitted.  Permission from IEEE must be obtained for all other uses, in any current or future media, including reprinting/republishing this material for advertising or promotional purposes, creating new collective works, for resale or redistribution to servers or lists, or reuse of any copyrighted component of this work in other works.}
\newcommand\copyrightnotice{%
\begin{tikzpicture}[remember picture,overlay]
\node[anchor=south,yshift=10pt] at (current page.south) {\fbox{\parbox{\dimexpr\textwidth-\fboxsep-\fboxrule\relax}{\copyrighttext}}};
\end{tikzpicture}%
}

\PassOptionsToPackage{capitalise,noabbrev}{cleveref}
\usepackage{hyperref}
\hypersetup{
    colorlinks=false,
    linkcolor=blue,
    urlcolor=blue,
}
\usepackage{cleveref}

\newcommand{\ft}{\textsc{FineTune}\xspace}
\newcommand{\tstep}{\textsc{TrainStep}\xspace}
\newcommand{\eval}{\textsc{Evaluate}\xspace}

\newcommand{\valsub}{\textsc{ValSub}\xspace}
\newcommand{\mae}{\textsc{MAE}\xspace}
\newcommand{\mse}{\textsc{MSE}\xspace}
\newcommand{\cossim}{\textsc{Cos}\xspace}
\newcommand{\canbdist}{\textsc{Canb}\xspace}
\newcommand{\corr}{\textsc{Corr}\xspace}
\newcommand{\allcorr}{\textsc{AllC}\xspace}
\newcommand{\randone}{\textsc{Rand\#1}\xspace}
\newcommand{\randtwo}{\textsc{Rand\#2}\xspace}
\newcommand{\randthree}{\textsc{Rand\#3}\xspace}
\newcommand{\oracle}{\textsc{Oracle}\xspace}

\newcommand{\linebreakand}{%
  \end{@IEEEauthorhalign}
  \hfill\mbox{}\par
  \mbox{}\hfill\begin{@IEEEauthorhalign}
}

\begin{document}

\title{ValSub: Subsampling Validation Data to Mitigate Forgetting during ASR Personalization}

\author{\IEEEauthorblockN{Haaris Mehmood\IEEEauthorrefmark{1}, Karthikeyan Saravanan\IEEEauthorrefmark{1}, Pablo Peso Parada\IEEEauthorrefmark{1}, David Tuckey\IEEEauthorrefmark{1},\\
Mete Ozay\IEEEauthorrefmark{1}, Gil Ho Lee\IEEEauthorrefmark{2}, Jungin Lee\IEEEauthorrefmark{2}, and Seokyeong Jung\IEEEauthorrefmark{2}}
\IEEEauthorblockA{\IEEEauthorrefmark{1}Samsung R\&D Institute UK (SRUK), \IEEEauthorrefmark{2}Samsung Electronics, South Korea
}}

\maketitle

\copyrightnotice

\begin{abstract}
Automatic Speech Recognition (ASR) is widely used within consumer devices such as mobile phones. Recently,  personalization or on-device model fine-tuning has shown that adaptation of ASR models towards target user speech improves their performance over rare words or accented speech. Despite these gains, fine-tuning on user data (target domain) risks the personalized model to forget knowledge about its original training distribution (source domain) i.e. catastrophic forgetting, leading to subpar general ASR performance. A simple and efficient approach to combat catastrophic forgetting is to measure forgetting via a validation set that represents the source domain distribution. However, such validation sets are large and impractical for mobile devices. Towards this, we propose a novel method to subsample a substantially large validation set into a smaller one while maintaining the ability to estimate forgetting. We demonstrate the efficacy of such a dataset in mitigating forgetting by utilizing it to dynamically determine the number of ideal fine-tuning epochs. When measuring the deviations in per user fine-tuning epochs against a 50x larger validation set (oracle), our method achieves a lower mean-absolute-error (3.39) compared to randomly selected subsets of the same size (3.78-8.65). Unlike random baselines, our method consistently tracks the oracle's behaviour across three different forgetting thresholds.
\end{abstract}


\section{Introduction}
\vspace{-0.1cm}
Automatic Speech Recognition (ASR) models have been greatly optimized in recent years to run within the memory and computational constraints of devices such as mobile phones \cite{DBLP:journals/taslp/PrabhavalkarHSSW24}. This has led to the development of data-private end-to-end ASR models, enabling use-cases such as on-device voice assistants, translators, speech transcribers, etc. Despite several improvements, modern ASR systems still show poor performance when transcribing under adverse acoustic conditions (e.g. background noise, babble noise or reverberation) \cite{DBLP:conf/interspeech/TsunooSNK021, DBLP:conf/interspeech/ParadaDSO22}, disordered speech \cite{DBLP:journals/corr/abs-2106-10259}, heavy accents \cite{DBLP:conf/icassp/ShiYLLFWQX21} and rare-words \cite{DBLP:conf/asru/JalalPPMSKZDLLJ23}. 

Recently, personalization \cite{DBLP:journals/corr/abs-2310-09988} has been shown to improve ASR performance, specifically for rare-words and heavy accents, by fine-tuning ASR models towards the end-user \cite{DBLP:conf/interspeech/SimZB19, DBLP:conf/emnlp/TomanekZPVB21, DBLP:conf/interspeech/HuoHSGMSSB22}. This paradigm typically involves privacy preserving fine-tuning of ASR models on user-data, operating entirely on-device.
Several prior works solve specific problems with ASR personalization such as the lack of annotated data \cite{DBLP:conf/asru/ZhangRMTPJSLLJ23, DBLP:conf/icassp/ZhaoLZCG18}, fine-tuning sample selection \cite{DBLP:conf/acl/KothawadeMHKIRJ23}, accented speech adaptation \cite{DBLP:journals/corr/abs-2307-00453}, continual adaptation \cite{DBLP:conf/eusipco/Eeckth22} and parameter efficient adaptation \cite{DBLP:conf/emnlp/TomanekZPVB21}. An alternative approach to fine-tuning is contextual biasing which incorporates a new block in the ASR pipeline in order to improve rare-word recognition \cite{DBLP:conf/slt/MunkhdalaiWPSLRS22}.

\begin{figure}[t]
  \centering
  \includegraphics[width=\columnwidth]{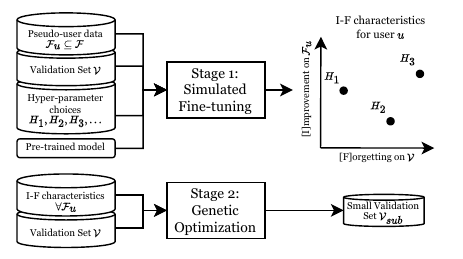}
  \vspace{-0.6cm}
  \caption{A high-level overview of our proposed method -- Importance based Subsampling of Validation Data.
  }
  \label{fig:pipeline}
  \vspace{-0.6cm}
\end{figure}

Practical use and deployment of personalization is affected by several considerations such as poor audio quality (e.g., with noise) or issues with data collection (e.g., incomplete recordings); bolstering the need for an effective on-device mechnanism to validate the performance of a personalized model. Furthermore, fine-tuning models on unseen distributions of data leads to catastrophic forgetting \cite{DBLP:conf/interspeech/LikhomanenkoXPT21}. Prior methods have proposed solutions \cite{doi:10.1073/pnas.1611835114, DBLP:conf/interspeech/SimZB19, DBLP:conf/asru/SimJMZBBGKKLZZ19, DBLP:conf/eusipco/Eeckth22, DBLP:conf/emnlp/TomanekZPVB21} to reduce the impact of catastrophic forgetting but they do not eliminate the problem. All such methods require back-propagating additional losses during training which limit their feasibility for on-device deployment due to significantly higher memory requirements. An efficient solution is to validate the personalized model over its source validation set: the model is rejected if prediction error on that set increases by a pre-determined threshold. However, this solution is still challenging for on-device deployment since typical validation sets would be too large to fit and evaluate on edge devices.
\vspace{0.5em}
\\
The main contributions of our work are as follows:

\begin{itemize}

\item A novel subsampling method to \textbf{produce a substantially smaller validation set that matches the behaviour of a large validation set} in measuring forgetting of previous knowledge during personalization. Our simple, derivative-free optimization approach is applicable to any off-the-shelf model and specifically does not require prior access to any personalization statistics or end-user data during the selection process. We are able to produce subsets up to 50x smaller than the original whilst maintaining comparable performances.
\item An investigation into the effectiveness of our subsampling method. We make use of an early stopping algorithm and our small validation set to \textbf{dynamically determine the ideal fine-tuning epochs for a given forgetting threshold}. We then measure how closely our small validation set can match the performance of the full validation in determining the ideal fine-tuning epochs.

\end{itemize}

Fig.~\ref{fig:pipeline} describes the proposed method of selecting a \textit{small} validation set through importance-based subsampling. The first stage in this process involves obtaining I-F characteristics of a target model using only $\mathcal{S}$. The next stage involves a derivative-free genetic optimization process to reduce a large validation set $\mathcal{V}$ into a small subset $\mathcal{V}_{sub}$, utilizing the previously generated I-F characteristics. Finally, to evaluate its effectiveness, we use $\mathcal{V}_{sub}$ to limit forgetting when training on user data, the process described under Algorithm \ref{alg:ft-es}.

\section{Background}
Derivative-free optimization methods \cite{DBLP:journals/jgo/RiosS13} can be employed in scenarios where the computation of the objective function's derivatives is either inefficient or simply cannot be performed. Genetic algorithms \cite{DBLP:journals/mta/KatochCK21} take inspiration from evolutionary biology to iteratively optimize an objective function without computing its derivatives.

Word error rate reduction (WERR) is a popular choice in ASR literature \cite{DBLP:journals/corr/abs-2307-00453} to measure the change in performance when fine-tuning a pre-trained model. WERR generally scales from 0\% (no change in error) to 100\% (all errors removed) and is a function to measure the amount of relative improvement in the error rate of an ASR model. Word error rate (WER) is defined as the minimum normalized edit distance function between the ground truth and predicted transcripts.

\vspace{0.25em}
\section{Proposed Approach}
\label{proposed_approach}
\subsection{Setup}
This section describes the proposed method: importance subsampling of validation data (\valsub). We assume a scenario where speech recognition models are privately fine-tuned on edge-devices with the objective to improve under-performing speech scenarios such as rare accents, rare words or disordered speech~\cite{ DBLP:journals/corr/abs-2106-10259, DBLP:conf/interspeech/SimZB19,DBLP:conf/interspeech/SimCGCMB21}. 

Let $\mathcal{S} = \{(\mathbf{x}_i, \mathbf{y}_i) : 1 \le i \le | \mathcal{S} |\}$ denote source domain (pre-training) and ${\mathcal{T} = \{(\mathbf{x}_i, \mathbf{y}_i) : 1 \le i \le | \mathcal{T} |\}}$ denote target domain (user data) datasets. Both datasets contain samples of paired audio recording $\mathbf{x}$ and transcript $\mathbf{y}$. To maintain privacy of end-users, we assume that $\mathcal{T}$ is not made available in the same environment as $\mathcal{S}$, that is, $\mathcal{T} \cap \mathcal{S} = \varnothing$ and $| \mathcal{T} | \neq | \mathcal{S} | $.

Personalization (or fine-tuning) implies optimizing a pre-trained ASR model (parameterized by $\theta$) over a subset of some target domain $\mathcal{T}$ to improve its performance on $\mathcal{T}$. Typically during this process, the model's performance on the source domain $\mathcal{S}$, deteriorates, since the model forgets previous knowledge about $\mathcal{S}$ \cite{DBLP:conf/asru/SimJMZBBGKKLZZ19}.
 
 With personalization, arbitrarily choosing hyper-parameters, especially the number of training epochs, could lead to either fine-tuning too little, (low improvement over $\mathcal{T}$) or too much (high forgetting over $\mathcal{S}$). In the extreme, high forgetting could mean the model loses its generalization capability and overfits to the user data, thereby failing to transcribe any speech other than the limited and specific data used for personalization. 
 It remains an open challenge to be able to identify optimal hyper-parameters for model fine-tuning without prior access to $\mathcal{T}$.

A simple solution to this is to modify the standard training procedure to periodically measure the deterioration over the source domain validation set and to early stop training when this deterioration exceeds a pre-defined threshold. 
Algorithm \ref{alg:ft-es} is an example of a fine-tuning algorithm with early stopping. After every $|\mathcal{B}|$ batches of training steps ($\tstep$), the source-domain forgetting is evaluated using a validation set $\mathcal{E}$ (\eval). If this score ($\gamma$) exceeds a pre-defined threshold ($\Gamma$) or the model has been trained for the maximum number of epochs ($\Omega$), fine-tuning stops and the last within-threshold ($\gamma \le \Gamma$) model weights are returned. 

The source domain validation data $\mathcal{E}$ is a key component of Algorithm \ref{alg:ft-es}. Our aim is to make this dataset as small as possible to make it feasible for memory constrained mobile devices, all the while still retaining important characteristics.

Towards this objective, we define \textit{improvement-forgetting} (I-F) characteristics as follows. (A) Improvement denotes the relative improvement of a model's performance on user data after fine-tuning. (B) Forgetting implies the relative deterioration of a model's performance on source data after fine-tuning. Next, we describe how I-F is used to produce a small scale validation set.

\begin{figure}[t]
\vspace{-1em}
{
\begin{algorithm}[H]
\caption{$\pmb{\ft}(\theta, \mathcal{D}, \mathcal{E}, \Gamma, \Omega)$ with early stopping.} 
\label{alg:ft-es}
  \begin{algorithmic}
  \small
    \State Input: $\theta$ (pre-trained model), $\mathcal{D}$ (local user data),\\$\mathcal{E}$ (validation data), $\Gamma$ (forgetting threshold), $\Omega$ (maximum number of epochs)
    \State Initialize: $\gamma = 0$ (forgetting score), $\omega=0$ (epoch), $\theta_{0} = \theta$
    \While{$\gamma \le \Gamma$ \textbf{and} $\omega \le \Omega$}
        \State $\theta_{1} = \theta_{0}$
        \For{$b \in \mathcal{B} \sim \mathcal{D}$}
            \State $\theta_{0} = \tstep(\theta_{0}, b)$ \Comment{train batch}
        \EndFor
        \State $\gamma = \eval(\theta_{0}, \mathcal{E})$ \Comment{record forgetting score}
        \State $\omega = \omega + 1$
    \EndWhile
    \State Return: $\theta_{1}$
\end{algorithmic}
\end{algorithm}
}
\vspace{-1.75em}
\end{figure}

\subsection{Importance based subsampling of validation data}
\vspace{-0.125em}
Towards this, we assume the source domain data $\mathcal{S}$ is split into 3 disjoint randomly sampled subsets: $\mathcal{P} \cup \mathcal{V} \cup \mathcal{F} = \mathcal{S}$. $\mathcal{P}$ is the data used during model pre-training. $\mathcal{F}$ is proxy user-data used to \textit{simulate fine-tuning} and gather I-F characteristics. Lastly, $\mathcal{V}$ is the data used for validation. Let $\mathcal{Y}$ be defined as all the ground truth transcripts in $\mathcal{V}$. The small-scale validation dataset is subsampled from $\mathcal{V}$.

Personalization data, typically only belonging to a single user, is substantially small relative to $\mathcal{P}$ and naturally exhibits high intra-set similarity e.g. speaker accent or transcript distribution \cite{userlibri}. Considering this, we assume that $U$ pseudo-user datasets (indexed by $u$) can be extracted from $\mathcal{F}$ by 
$\mathcal{F}_{1,\ldots,U} = \{ \mathcal{F}_u \subseteq \mathcal{F} : 1 \le u \le U \}$, with each $\mathcal{F}_{u}$ exhibiting some notion of intra-user similarity.

To maintain user privacy and practicality for deployment, our methodology assumes no access to target domain user data $\mathcal{T}$ for either of the two stages. We hypothesize that genetic optimization using I-F characteristics of pseudo-users arising from $\mathcal{S}$ should be sufficient to capture fine-tuning behaviour using $\mathcal{T}$. We empirically verify this in the experiments section by measuring the effectiveness of the small-scale validation sets in predicting ideal fine-tuning epochs.

The pre-trained model is evaluated using $\mathcal{V}$ before the first stage and the predictions, denoted by $\mathcal{Y}_0$, are stored for later use.

During the {\bf first stage}, fine-tuning outcomes (I-F characteristics) for $U$ pseudo-user datasets $\mathcal{F}_{1,\ldots,U}$ against a pre-determined selection of $K$ hyper-parameters $\{H_1, H_2, \ldots, H_K\}$ are captured (see Fig.~\ref{fig:pipeline}). To calculate the improvement score (I), predictions of inputs in $\mathcal{F}_{u}$ before and after fine-tuning are required. Similarly, to calculate the forgetting score (F), $\mathcal{Y}_0$ and predictions of inputs in $\mathcal{V}$ after fine-tuning are required. $\mathcal{Y}_{1,\ldots,U}^{1,\ldots,K}$ denotes the validation set predictions for all $U$ users and $K$ hyper-parameters. The function to calculate the improvement and forgetting scores in this work are defined under Section \ref{sec-g-h}.

The {\bf second stage} involves a derivative-free optimization process. Let samples in $\mathcal{V}$ be indexed by the vector $\mathbf{I} = \{ 1, \ldots, |\mathcal{V}| \}$. The goal of the optimization process is to determine the optimal subset of indices $\mathbf{I}^{*}_{sub} \subset \mathbf{I}$ such that $\mathcal{V}_{\mathbf{I}^{*}_{sub}} = \{ v_i : \forall i \in \mathbf{I}^{*}_{sub} \}$ maximally mimics the forgetting characteristics captured by $K$ hyper-parameter choices and $U$ pseudo-users. We denote $\mathcal{V}_{\mathbf{I}^{*}_{sub}}$ as $\mathcal{V}_{sub}$ for readability.

A scoring function $g : \mathbb{R}^{M} \times \mathbb{R}^{M} \to \mathbb{R}$ returns the forgetting score for a vector of M predictions and ground truth transcripts. A utility function $h : \mathbb{R}^{N} \times \mathbb{R}^{N} \to \mathbb{R}$ measures how well $\mathcal{V}_{sub}$  mimics $\mathcal{V}$ by computing the pairwise similarity of $N = KU$ forgetting scores. There are several valid choices for $g$ and $h$, our choices in this paper are defined in Section \ref{sec-g-h}.

Since $\mathcal{I}_{sub} \subset \mathcal{I}$, evaluations of all samples in $\mathcal{V}$ can be stored and retrieved for every realization of $\mathcal{I}_{sub}$ without having to run inference on the per user models again making the optimization procedure very efficient. The optimization process can run for a desired number of iterations or until convergence.

Finally, $\mathbf{I}_{sub}$ is evaluated on the target distribution $\mathcal{T}$ i.e.\linebreak$\mathcal{T}_{1,\ldots,V} = \{ \mathcal{T}_v \subseteq \mathcal{T} : 1 \le v \le V \}$. First, the entire set of validation samples (i.e. $\mathcal{V}_\mathbf{I}$) and a pre-defined forgetting threshold ($\Gamma$) is used to fine-tune on data from $\mathcal{T}$ using Algorithm \ref{alg:ft-es}. Next, the results are used to calculate I-F scores (see Section \ref{sec-g-h}). These set of results form \oracle. The efficacy of $\mathcal{V}_{\mathbf{I}_{sub}}$ can then be determined either visually or analytically by measuring the distance between I-F scores produced using various realizations of $\mathcal{V}_{\mathbf{I}_{sub}}$ versus \oracle.

\section{Experimental Setup}
\label{exp_setup}
\subsection{Model}

This work uses a pre-trained E2E ASR model following the Conformer-Transducer architecture described in \cite{DBLP:conf/icassp/NarayananSPYCPV21}. Based on~\cite{DBLP:conf/interspeech/SimZB19}, we freeze all but the last three conformer encoder layers and all batch norm parameters \cite{DBLP:journals/ijcv/WuH20}. 

\subsection{Datasets}

\textbf{Pre-training}: The training dataset $\mathcal{P} \subset \mathcal{S}$ consists of 20k hours of in-house and public spoken English audio with ground truth transcripts arising from a variety of domains: telephony, LibriSpeech, far-field, search. We choose an in-house dataset due to the greater diversity of speakers and a realistic range of domains covered. The validation data $\mathcal{V} \subset \mathcal{S}\setminus\mathcal{P}$  has 5k randomly sampled utterances (6 hours of audio) which is used to evaluate the model's performance on held-out data.

\textbf{Simulated fine-tuning}: The combined fine-tuning dataset from the source domain $\mathcal{F} = S \setminus (\mathcal{P} \cup \mathcal{V})$ also consist of 5k utterances randomly sampled from $\mathcal{S}$. To procure per user fine-tuning sets $\mathcal{F}_{1,\ldots,U} \subseteq \mathcal{F}$ without having access to user identifiers, we make use of an off-the-shelf state-of-the-art (SOTA) speaker identification (SID) model \cite{DBLP:conf/interspeech/DesplanquesTD20}.
We project all utterances in $\mathcal{F}$ to the SID embedding space and then make use of auxiliary user centroids in the same space to assign each sample in $\mathcal{F}$ to a particular speech style. Auxiliary user centroids are generated by computing a mean of in-house human annotated data, separate from $\mathcal{S}$ or $\mathcal{T}$ and projected to the SID space, These samples are only used to supervise the clustering process for samples in $\mathcal{F}$ but not during pre-training or fine-tuning.

After clustering all samples in $\mathcal{F}$, the top-150 nearest samples per centroid are extracted with 12 centroids in total. As a form of cross-validation, we employ repeated random sub-sampling \cite{DBLP:journals/midm/Shan22}: from each such user, we randomly sample six sets of 24 samples each to give a total of $U=72$ `pseudo-users'. As a result, the per user sets $\mathcal{F}_{1,\ldots,72}$ inhibit high intra-set similarity in terms of speaker characteristics.

\begin{figure}[t]
\centering
\input{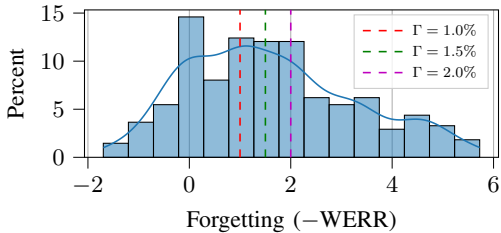}
\vspace{-0.25cm}
\caption{Forgetting histogram and density estimation after evaluating $\mathcal{V}$ on fine-tuned models of $U=72$ pseudo-users for $K=4$ hyper-parameters totalling 288 runs; 15 outlier runs ($<-6\%$ WERR) were removed. The mean (standard deviation) is 1.6 (1.7).}
 \captionsetup{font={small, it}}
 \label{fig:plot_distrib_pseudo_users}
 \vspace{-1.5em}
\end{figure}

\textbf{Target fine-tuning}: To evaluate the feasibility of our method, we require a fine-tuning dataset arising from a  distribution different from $\mathcal{S}$. For our experiments, we evaluate our method using an in-house dataset $\mathcal{T}$ consisting of 7 users belonging to the same region. We chose a region on which the performance of the pre-trained model is lower than the average of $\mathcal{F}$ (i.e. higher WER) as this makes the model more likely to forget the source domain when it is fine-tuned based on our internal experiments. Each target user data $\mathcal{T}_v$ consists of an equal split of mobile application launch/install commands (Apps) and mobile contact call/text commands (Contacts). The average utterance lengths and representative examples for these sets follow a similar format as used in \cite{DBLP:conf/asru/ZhangRMTPJSLLJ23}. Each target user has a total of 100 samples. We chose a different number of samples for the target users to demonstrate robustness of our technique to change in training samples between source and target fine-tuning datasets.

\textbf{Statistics}: The mean (standard deviation) WER for $\mathcal{F}_{1..U}$ is 8.55 (4.87), and for $\mathcal{T}_{1,\ldots,V}$ is 46.3 (8.71). The WER for $\mathcal{V}$ is 9.43.

\subsection{Choices for the scoring ($g$) and utility ($h$) functions}
\label{sec-g-h}
The \textbf{scoring function $g$} outputting the forgetting score is chosen as negative word error rate reduction ($-$WERR) for the  purpose of gathering data during stages 1 and 2.

For a given user $u$ and hyper-parameter $k$, defining $\text{WER}_0 = d(\mathcal{Y}_0,\, \mathcal{Y})$ and $\text{WER}_u^k = d(\mathcal{Y}_{u}^{k},\, \mathcal{Y})$, where $d$ denotes the minimum normalized edit distance function between transcripts, the forgetting score is defined as: 
\vspace{-0.12em}
\begin{equation}
    -\text{WERR}_u^k = (\text{WER}_u^k - \text{WER}_0)/\text{WER}_0
\end{equation}

Conversely, the function outputting the improvement score is defined as word error rate reduction (WERR).

For the \textbf{utility function $h$}, which pairwise compares forgetting on different subsets of $\mathcal{V}$, we evaluate several similarity and distance functions. Specifically, we experiment with negative mean squared error (\mse), negative mean absolute error (\mae), cosine similarity (\cossim), Pearson correlation coefficient (\corr), and negative Canberra distance (\canbdist) \cite{DBLP:journals/corr/abs-2307-02694, journals/plosone/simmetrics, DBLP:journals/pr/BlancoMalloMRB23}.

\subsection{Choices for \texorpdfstring{$\Gamma$}{Gamma},  \texorpdfstring{$\Omega$}{Omega} and \texorpdfstring{$k$}{k}}

We target three forgetting thresholds $\Gamma \in \{ 1.0\%, 1.5\%, 2.0\% \}$ for the validation set $\mathcal{V}$ to evaluate the robustness of the proposed method against changing thresholds. To gather I-F characteristics with simulated fine-tuning, we use a collection of parameters that lead to good support (high number of runs) near the threshold regions. Let $k$ be a 2-tuple describing epochs to train and the learning rate to use respectively, that is, ${k \in \{ (8, 1\text{e-}5), (16, 1\text{e-}5), (16, 5\text{e-}6), (32, 5\text{e-}6) \}}$. We chose epochs and learning rate as the hyper-parameter to vary as it naturally leads to greater improvement but also greater forgetting when increased. Fig. \ref{fig:plot_distrib_pseudo_users} shows a distribution of forgetting scores generated using $\mathcal{Y}_{1,\ldots,U}^{1,\ldots,K}$, i.e., 
$-$WERR scores obtained after training of all $U$ users and $K$ hyper-parameter choices. After 15 epochs, 97\% of psuedo-users have all correct predictions and hence we set $\Omega = 15$ as maximum number of epochs for stage 3.

\subsection{Genetic optimization}
For the genetic optimization procedure \cite{DBLP:journals/mta/KatochCK21}, each ``gene” takes any unique value from a set of sample IDs in $\mathbf{I}$. For the first generation, 20 candidates for $I_{sub}$ are generated, and each generation has 100 IDs uniformly sampled at random (u.s.r) as genes. For each generation, the top-10 scoring candidates (based on $h$) are selected as parents. Cross-over operations produce 10 offsprings (the first half of first parent is combined with the second half of the second parent and so on). 10\% of all genes in each offspring are then ``mutated” to a different ID u.s.r. The procedure is terminated after 650 generations.

\vspace{-0.375em}
\subsection{Baselines}
\vspace{-0.125em}
We compare various choices for $h$ against two types of baselines. (1) \randone, \randtwo \& \randthree: selecting $3 \times \mathbf{I}_{sub}$ randomly from a uniform distribution. (2) \allcorr: selecting $\mathbf{I}_{sub}$ randomly from samples that have no errors in $\mathcal{Y}_0$; a method  based on \cite{DBLP:conf/interspeech/SimCGCMB21}.

\vspace{0.125em}
\section{Results and Analyses}
\label{results}
\subsection{Choice of utility function \texorpdfstring{$h$}{h}}
\begin{figure}[t]
 \centering
 \resizebox{0.8\columnwidth}{!}{%
    \includegraphics{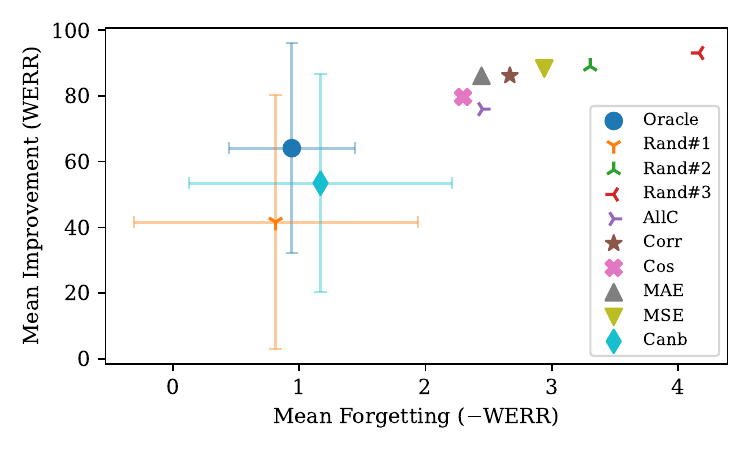}
 }
\vspace{-0.75em}
 \caption{ An I-F plot for mean values of $V=42$ users using a target threshold of $\Gamma=1.5\%$ for forgetting. Error bars denoting standard deviation are only shown for \oracle, \randone and \canbdist for brevity. Our proposed method using \canbdist is closest to the oracle.}
 \label{fig:plot_if}
\vspace{-1em}
\end{figure}
Fig. \ref{fig:plot_if} illustrates mean I-F scores using different choices for the utility function $h$. Performance of methods on three randomly generated validation subsets, an all correct subset (\allcorr), and an oracle score (\oracle) is also shown. We find that \randone is competitive to several of the evaluated methods. However, \canbdist is relatively closest to the oracle. This suggests that some form of normalization with respect to the size of the forgetting score vector helps in achieving better performance as it limits skewing. We compare relative distance since WERR can be large for small data sets, and vice-versa.

Table \ref{tab:abs-error} describes the pairwise mean-absolute-error (MAE) between \oracle, random baselines, \allcorr, and the proposed method with \canbdist for three different target thresholds and the average ($\Bar{\Gamma}$). The error is measured for the epochs trained per target user $v$ and then averaged. While the proposed method using \canbdist performs the best on average, \randone is best for $\Gamma=1.0$ because it tends to skew towards the minimum which coincides with \oracle in this case.

\begin{table}[hb]
  \caption{Pairwise mean (standard deviation) absolute error (MAE) in epochs trained when using \oracle, random baselines, an all correct baseline (\allcorr), or our proposed method (using \canbdist); evaluated at three different forgetting thresholds ($\Gamma$).}\label{tab:abs-error}
  \vspace{-0.25em}
  \centering
  \resizebox{0.8\columnwidth}{!}{%
  \begin{tabular}{@{}ccccc@{}}
  \toprule
    \multirow{2}{*}[-1.5pt]{\textbf{Method}}  & \multicolumn{4}{c}{\textbf{Pairwise MAE in Epochs Trained}} \\
    \cmidrule{2-5}
    & \textbf{$\Gamma=1.0\%$} & \textbf{$\Gamma=1.5\%$} & \textbf{$\Gamma=2.0\%$} & \textbf{$\Bar{\Gamma}$} \\
    \midrule
    \randone  & 2.71 (3.01)         & 4.07 (3.35)           & 4.57 (3.42)            &  3.78            \\
    \randtwo  & 8.17 (4.67)         & 6.79 (4.35)           & 5.86 (3.92)            &  6.94                  \\
    \randthree  & 10.36 (3.74)         & 8.88 (4.18)        & 6.71 (4.28)            &  8.65 \\
    \allcorr & 5.86 (4.89)            & 5.29 (3.98)         & 5.02 (3.86)            &  5.39 \\
    \canbdist   & 2.86 (2.67)   & 3.83 (3.02)   & 3.48 (3.15)      & \textbf{3.39} \\
    \bottomrule
  \end{tabular}
  }
\end{table}

\begin{figure}[ht]
\vspace{-0.75em}
 \centering
 \resizebox{0.68\columnwidth}{!}{%
 \includegraphics{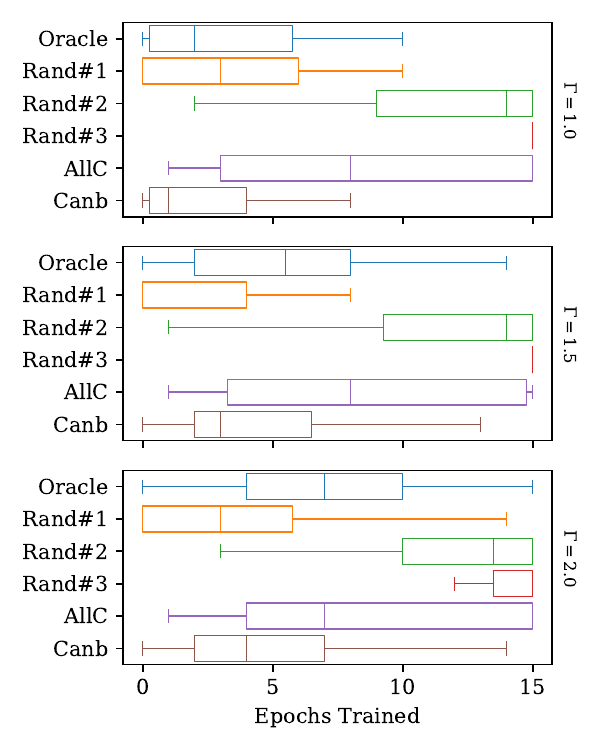}
 }
 \vspace{-0.75em}
 \caption{Box and whiskers plot of epochs trained for oracle, baselines and the proposed method. Random baselines maintain a skewed distribution towards either the minimum or maximum epochs. The all correct baseline maintains a static uniform-like distribution. \textbf{The proposed method closely follows oracle across three thresholds}. }
 \label{fig:plot_ep_distrib}
 \vspace{-0.25em}
\end{figure}

Fig.~\ref{fig:plot_ep_distrib} illustrates the distribution of epochs trained using target user data $\mathcal{T}$. We compare all of the baseline methods and our best method against \oracle. While \randone has a median at 0 (min epochs), \randtwo and \randthree have a median at or near 15 (max epochs). Thus, using random subsets is not a feasible alternative. \allcorr has a distribution very different from \oracle. On the contrary, our method with the lowest MAE (\canbdist) has a distribution very close to \oracle. This positively affirms our hypothesis that using only data from the source domain $\mathcal{S}$, we are able to produce a small validation set which closely follows the behaviour of a much larger validation set when fine-tuning using target domain $\mathcal{T}$ data.

\subsection{Effect of Validation Subset Size}

We investigate the number of data samples required for a validation subset. 50, 100 and 1000 samples give good results with the 50 samples experiment surprisingly performing the best as shown in Table \ref{tab:rmse-size}. We conjecture that a very small number (10) of samples could limit expressiveness of a dataset to effectively capture forgetting behaviour and, at the same time, a large subset (500) might require more cross-over and mutation operations per round.
\begin{table}[hb!]
  \caption{\small Mean absolute error (MAE) scores with varying subset sizes for $\Gamma = 1.5\%$ and \canbdist as the utility function.}
  \vspace{-0.25em}
  \label{tab:rmse-size}
  \centering
  \resizebox{0.70\columnwidth}{!}{%
  \begin{tabular}{rcccccc}
    \toprule
    \textbf{Subset Size} & 10 & 50 & 100 & 500 & 1000 \\
    \midrule
    \textbf{MAE} & 4.38 & 2.21 & 3.83 & 7.88 & 3.76  \\    \bottomrule
  \end{tabular}
  }
\end{table}

\section{Conclusion}
ASR personalization towards a target domain typically leads to loss of previously acquired source domain knowledge. An approach to mitigate forgetting is to measure and limit it using source domain validation sets. We have presented a method to subsample from a large validation set via similarity matching of forgetting scores between the subsampled and the original set to produce a 50x smaller subset. We demonstrated the efficacy of our method by using the subsampled dataset to optimally determine the number of fine-tuning epochs for users arising from a separate, target distribution. Empirical evaluations show that our method outperforms random selection and an all correct random selection at three different target forgetting thresholds. In future, we plan to investigate further choices for the utility function used during optimization as well as evaluate on additional forgetting thresholds.

\clearpage
\IEEEtriggeratref{15}
\bibliographystyle{ieeetr}
\bibliography{refs}

\end{document}